\documentclass[conference]{IEEEtran}
\IEEEoverridecommandlockouts
\usepackage[T1]{fontenc}
\usepackage{cite}
\usepackage{amsmath,amssymb,amsfonts}
\usepackage{algorithmic}
\usepackage{graphicx}
\graphicspath{{walkthrough_assets/}{./}}
\usepackage{textcomp}
\usepackage{xcolor}
\usepackage{url}
\usepackage{array}
\usepackage{enumitem}
\usepackage{listings}
\usepackage{tcolorbox}
\usepackage{booktabs}
\usepackage{tikz}
\usetikzlibrary{arrows.meta,positioning,fit,backgrounds,calc}

\lstdefinelanguage{Rules}{
  morekeywords={
    id, severity, pattern, title, rationale, asset_category,
    code_rules, log_rules, topology_rules
  },
  sensitive=true,
  morecomment=[l]{\#},
  morestring=[b]"
}

\lstdefinestyle{modelstyle}{
  basicstyle=\ttfamily\scriptsize,
  breaklines=true,
  breakatwhitespace=false,
  numbers=left,
  numberstyle=\tiny,
  stepnumber=1,
  numbersep=5pt,
  frame=single,
  columns=fullflexible,
  keepspaces=true,
  showstringspaces=false,
  captionpos=b
}

\tcbuselibrary{skins,breakable}

\newcommand{\promptfontsize}{\footnotesize}

\newtcolorbox{promptbox}[1][]{
  breakable,
  enhanced,
  colback=gray!5,
  colframe=gray!40,
  fontupper=\ttfamily\promptfontsize,
  fonttitle=\bfseries\promptfontsize,
  title=#1,
  left=1em, right=1em, top=1em, bottom=1em,
  sharp corners
}

\definecolor{detblue}{RGB}{31,86,140}
\definecolor{decpurple}{RGB}{104,66,150}
\definecolor{grdgreen}{RGB}{40,110,70}
\definecolor{refred}{RGB}{150,45,45}
\definecolor{boxgray}{RGB}{90,90,90}

\tikzset{
  fignode/.style={draw, rounded corners=1.5pt, align=center, font=\scriptsize,
                  inner sep=3pt, minimum height=6mm},
  ndet/.style={fignode, draw=detblue,   fill=detblue!6},
  ndec/.style={fignode, draw=decpurple, fill=decpurple!7},
  ngrd/.style={fignode, draw=grdgreen,  fill=grdgreen!7},
  nio/.style={fignode,  draw=boxgray,   fill=black!4},
  nref/.style={fignode, draw=refred,    fill=refred!7},
  fflow/.style={-{Stealth[length=1.6mm]}, semithick, black!55},
  fgood/.style={-{Stealth[length=1.6mm]}, semithick, grdgreen},
  fbad/.style={-{Stealth[length=1.6mm]}, semithick, refred},
  freplan/.style={-{Stealth[length=1.6mm]}, semithick, densely dashed, orange!60!black},
}

\usepackage{hyperref}
\hypersetup{
  colorlinks=true,
  linkcolor=blue,
  citecolor=blue,
  urlcolor=blue,
  linktoc=all
}

\begin{document}

\title{CyberLLM: A Multi-Agent LLM Framework for Autonomous Detection and Guarded Response in Automotive Cybersecurity\\
\thanks{
This work has received funding from the European Chips Joint Undertaking under Framework Partnership Agreement No 101139789 (HAL4SDV) including the national funding from the Federal Ministry of Research, Technology and Space of Germany under grant number 16MEE00471K. The responsibility for the content of this publication lies with the authors.
}
}

\author{
\IEEEauthorblockN{
Nenad Petrovic, Oussama Jeddou, Feres Ben Fraj, \\ Vahid Zolfaghari, Fengjunjie Pan,  Andre Schamschurko, Alois Knoll
}

\IEEEauthorblockA{
\textit{Chair of Robotics, Artificial Intelligence and Real-Time Systems}\\
Technical University of Munich, Munich, Germany\\
Email: nenad.petrovic@tum.de, oussamajaddou@gmail.com, ge87qap@mytum.de, \\ v.zolfaghari@tum.de, f.pan@tum.de, andre.schamschurko@tum.de, k@tum.de
}
}
\maketitle

\begin{abstract}
Software-Defined Vehicles (SDVs) expand the automotive attack surface across source code, runtime logs, and deployment topologies, while safety constraints forbid autonomous agents from acting without oversight. This paper presents \emph{CyberLLM}, a multi-agent, LLM-orchestrated framework that autonomously detects vulnerabilities and executes remediations under a formal, runtime safety guard. Detection combines a deterministic layer (regex rules, AST analyzers, and topology graph checks) with an LLM refinement pass, so a high-recall floor is complemented by high-precision reasoning. A decision agent aggregates findings, tags them with a human-centric asset taxonomy, and selects a tiered response, ratcheting its confidence with signed cross-session memory and re-planning feedback. Every action is validated against four contextual security properties and an independent action-alignment oracle before it is allowed to run, and refused actions trigger escalation and re-planning. A symmetric attack pipeline generates and replays exploits so both sides can be exercised on the same scenarios. On an independent, ground-truthed benchmark of nine original automotive ECU modules in C, C++, and Rust seeding 47 layered vulnerabilities plus clean controls, the always-on deterministic layer covers 34\% of the labeled vulnerabilities at perfect precision, and adding the grounded LLM refinement and completeness passes roughly doubles coverage to about 70\% (F1 $0.83$) while producing zero false positives on the clean controls. The results indicate that LLM agents can perform useful autonomous cyber-defense when wrapped in a deterministic, auditable safety envelope.
\end{abstract}


\section{Introduction}

The transition toward Software-Defined Vehicles (SDVs) has moved a growing share of vehicle functionality into software that is updated over the air and composed from heterogeneous components, in-vehicle services, and increasingly LLM-based assistants. This shift dramatically enlarges the cyber-attack surface: vulnerabilities can originate in application source code, manifest as anomalies in telemetry and CAN logs, or be latent in the deployment topology that connects electronic control units (ECUs), gateways, and external agents~\cite{iso21434}. Traditional security tooling addresses these layers in isolation with rule-based scanners, signature-based log monitors, and manual configuration reviews, none of which reason across layers or act autonomously.

Large Language Models (LLMs) offer a compelling alternative: they ingest heterogeneous text artifacts (code, logs, configurations), reason over domain concepts, and can propose concrete remediations~\cite{pearce2023repair}. However, using LLM agents for cyber-defense raises two coupled challenges. First, LLM detection is \emph{probabilistic}: it may hallucinate findings or miss issues~\cite{ji2023hallucination}, which is unacceptable as the sole basis for a security decision. Second, an autonomous agent that can act on a safety-critical vehicle must never take an unsafe or unauthorized action, even when inputs are adversarially manipulated (e.g., via prompt injection or memory poisoning~\cite{greshake2023injection,owaspagentic}). Autonomy and safety are therefore in tension.

This paper addresses that tension with \emph{CyberLLM}, a multi-agent framework that pairs autonomous detection and response with a deterministic, auditable safety envelope. We tackle three challenges in particular: (i) achieving high detection recall without sacrificing precision, by combining deterministic analyzers with LLM refinement; (ii) making autonomous response decisions that are conservative and repeatable, by wrapping a probabilistic decision agent in an escalate-only policy informed by signed memory; and (iii) guaranteeing that no action executes unless it satisfies formal, contextual security properties, enforced at runtime by an independent guard.

The main contributions of this paper are:
\begin{itemize}[leftmargin=1.2em]
  \item A layered detection design that merges regex, AST-based (Bandit/Semgrep), and topology-graph analyzers with an LLM refinement pass into a single, deduplicated finding stream, where the deterministic floor grounds and constrains the LLM.
  \item A decision-and-response engine that selects a tiered mitigation, ratchets it with cross-session signed memory and re-planning feedback, and executes it through an MCP-style tool registry with per-step verification and rollback.
  \item A runtime security guard that enforces four contextual security properties plus an independent action-alignment oracle, so that a subverted scanner cannot cause an unsafe action to be committed.
  \item A symmetric attack pipeline that generates and replays exploits, enabling both defense and attack sides to be measured on identical scenarios, together with an evaluation on a hand-labeled automotive corpus.
\end{itemize}

The remainder of this paper is organized as follows. Section~\ref{sec:related} reviews the five works that inform our design and positions our contribution. Section~\ref{sec:impl} details the architecture and the defense and attack workflows. Section~\ref{sec:eval} describes the experimental setup and reports results. Section~\ref{sec:concl} concludes and outlines future work.

\section{Background and Related Works}
\label{sec:related}

CyberLLM is synthesized from five complementary lines of work; we summarize each, state precisely which elements we adopt, and describe how we go beyond them.

\textbf{Contextual agent security (safety envelope).} Siu \emph{et al.}~\cite{siu2026framework} formalize LLM-agent security around the insight that the \emph{same} action may be legitimate or malicious depending on the execution context \(C_t=(p, Tr_{t-1}, M_t, E_t, S_{\mathrm{auth},t}, G)\), and define four properties (task alignment, action alignment, source authorization, data isolation) together with oracle functions. We adopt these four properties directly as the runtime checks in our \texttt{SecurityGuard}, and we implement the high-value action-alignment oracle \(H_a\) as a \emph{separate} LLM judge that can only tighten, never loosen, the deterministic decision. Beyond the framework, we make provenance mandatory at ingestion, extend the checks to trajectory-level (compositional and temporal) safety, and apply source authorization to the memory channel.

\textbf{Human-centric threat taxonomy (attack surface).} Stappen \emph{et al.}~\cite{stappen2026agent2agent} (AgentHeLLM) separate \emph{what} is protected (a taxonomy of seven human-centric asset categories) from \emph{how} it is attacked (a graph of actors, datasources, and interaction primitives) for in-vehicle agent-to-agent settings. We adopt the seven-category asset taxonomy to tag findings and the actor/datasource graph model for our deployment scanner, and we reuse the attack-path perspective to structure our attack pipeline. Beyond the taxonomy, we operationalize it as executable topology rules and as memory-poisoning defenses.

\textbf{Model-driven safety and security by design (detection pattern).} Petrovic \emph{et al.}~\cite{petrovic2026fusa} propose LLM-empowered workflows combining metamodels, instance creation, OCL-style rules, and RAG context for automotive functional safety and security. We adopt their pattern of grounding the LLM with retrieved, machine-checkable rules rather than free generation, realizing it as a YAML rule base plus deterministic checkers whose hits become context for the LLM scanners. Beyond their code-analysis workflow, we generalize the pattern across three input modalities (code, logs, topology) and add AST-based analyzers as an intermediate layer.

\textbf{Schema-driven multi-agent execution (backbone).} Härer~\cite{harer2024multiagent} describes a schema-driven multi-agent LLM system with explicit prompt/data/action functions and stack-based execution, including cybersecurity test cases. We adopt this as the executable backbone: every agent owns an LLM client, a data dictionary that fills a prompt template, and an objective used by the guard. Beyond the backbone, we add delegation semantics in which a child agent inherits the objective but not the parent's trusted sources.

\textbf{Orchestrated edge defense (layered shape).} Collins \emph{et al.}~\cite{collins2026orchestrated} give a three-layer blueprint (edge agents, LLM orchestration, security-intelligence/learning) for adaptive cyber-defense at the edge, including a tiered response and an adaptive-learning loop. Tiered response is adopted (soft/moderate/hard) and the learning loop, realized as a signed incident store consulted by the decision agent. Beyond the blueprint, we make learning loop tamper-evident (HMAC-signed, source-verified reads) so that recalled precedent cannot be poisoned.

In contrast to these works, which each address one facet, CyberLLM integrates detection grounding, tiered guarded response, formal runtime enforcement, tamper-evident memory, and a symmetric attack simulator into a single end-to-end pipeline aimed at the automotive domain.

\section{Implementation Overview}
\label{sec:impl}

\subsection{System Architecture}
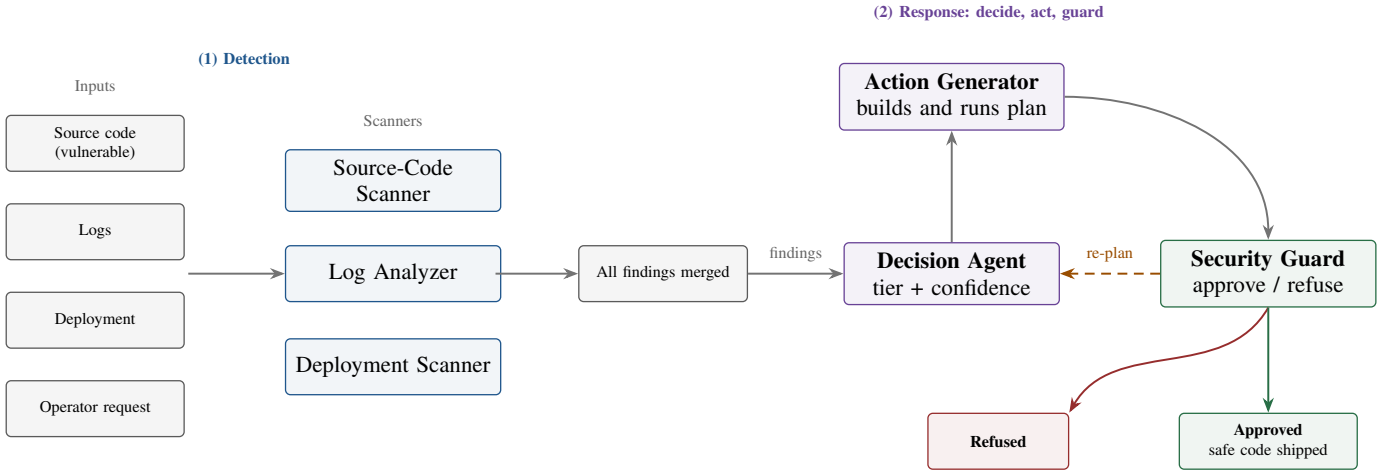
\begin{figure*}[t]
    \centering
    \resizebox{\textwidth}{!}{%
    \begin{tikzpicture}
      \node[font=\tiny\bfseries, detblue]   at (1.6,4.2) {(1) Detection};
      \node[font=\tiny\bfseries, decpurple] at (9.6,4.7) {(2) Response: decide, act, guard};
      \node[nio, font=\tiny, text width=1.7cm] (src)  at (0,3.3)  {Source code (vulnerable)};
      \node[nio, font=\tiny, text width=1.7cm] (logs) at (0,2.35) {Logs};
      \node[nio, font=\tiny, text width=1.7cm] (dep)  at (0,1.4)  {Deployment};
      \node[nio, font=\tiny, text width=1.7cm] (op)   at (0,0.45) {Operator request};
      \node[font=\tiny, black!60] at (0,3.9) {Inputs};
      \node[ndet, text width=2.1cm] (sc) at (3.2,2.9) {Source-Code Scanner};
      \node[ndet, text width=2.1cm] (la) at (3.2,1.9) {Log Analyzer};
      \node[ndet, text width=2.1cm] (ds) at (3.2,0.9) {Deployment Scanner};
      \node[font=\tiny, black!60] at (3.2,3.55) {Scanners};
      \node[nio, font=\tiny, text width=1.6cm] (mrg) at (6.1,1.9) {All findings merged};
      \node[ndec, text width=2.1cm] (dec) at (9.2,1.9) {\textbf{Decision Agent}\\ tier + confidence};
      \node[ndec, text width=2.2cm] (act) at (9.2,3.8) {\textbf{Action Generator}\\ builds and runs plan};
      \node[ngrd, text width=2.1cm] (grd) at (12.6,1.9) {\textbf{Security Guard}\\ approve / refuse};
      \node[nref, font=\tiny, text width=1.3cm] (ref) at (9.7,0.1)  {\textbf{Refused}};
      \node[ngrd, font=\tiny, text width=1.7cm] (app) at (12.6,0.1) {\textbf{Approved}\\ safe code shipped};
      \draw[fflow] (1.0,1.9) -- (la.west);
      \draw[fflow] (4.3,1.9) -- (mrg.west);
      \draw[fflow] (mrg.east) -- node[above, font=\tiny, black!60]{findings} (dec.west);
      \draw[fflow] (dec.north) -- (act.south);
      \draw[fflow] (act.east) to[out=0, in=90] (grd.north);
      \draw[freplan] (grd.west) -- node[above, font=\tiny]{re-plan} (dec.east);
      \draw[fbad]  (grd.south) to[out=-120, in=60] (ref.north east);
      \draw[fgood] (grd.south) -- (app.north);
    \end{tikzpicture}}
    \caption{CyberLLM defense pipeline. Three scanners produce a merged finding set that feeds the decision agent; the action generator's plan is validated step by step by the security guard, and refusals trigger re-planning.}
    \label{fig:defense}
\end{figure*}

CyberLLM is a pipeline of schema-driven agents. Every agent derives from a common \texttt{BaseAgent} that owns an LLM client, a data dictionary bound to a prompt template, an \emph{objective} string that the guard reasons about, and a mandatory-provenance input type: an artifact cannot enter the pipeline without a \texttt{source} and a \texttt{trust\_level} drawn from a fixed lattice (\texttt{authenticated}, \texttt{external}, \texttt{tool}, \texttt{memory}). All detectors emit a uniform \texttt{Finding} record (\texttt{rule\_id}, \texttt{severity}, \texttt{evidence}, \texttt{explanation}, \texttt{source\_agent}, \texttt{asset\_category}, \texttt{line}), which lets deterministic and LLM outputs merge seamlessly. The overall two-stage pipeline, \emph{detection} then \emph{guarded response}, is shown in Fig.~\ref{fig:defense}.

\subsection{Defense Workflow}
The defense mode runs the three scanners, then an adaptive decision--action--guard loop that re-plans on refusal.

\textbf{Detection: deterministic layer.} The always-on detector is a deterministic static analyzer that needs no LLM and yields a reproducible ground-truth floor, with one entry point per modality. \texttt{scan\_code\_static} applies the YAML \texttt{code\_rules} as multiline, case-insensitive regexes, recording a line number and a short code excerpt as evidence for each hit. \texttt{scan\_log\_static} applies hand-written temporal/statistical detectors keyed to \texttt{LOG-*} rules, e.g., $\geq 10$ failed authentications in 60\,s (brute force), a broadcast storm above 500\,packets/s, a DNS covert channel flagged when a query's Shannon entropy exceeds 3.0, and bulk exfiltration above 10\,MB. \texttt{scan\_deployment\_static} loads the deployment YAML into an actor/datasource graph (paper 2) and checks topology invariants \texttt{TOP-001} to \texttt{TOP-008}, such as an external actor holding a write edge into memory. Every hit is emitted as a \texttt{StaticFinding} with the same shape as an LLM \texttt{Finding}, so the layers merge without translation. On source code an optional AST layer runs Bandit (\texttt{.py}) or Semgrep (other languages)~\cite{bandit,semgrep} in a sandboxed subprocess (bounded timeout, no shell), mapping analyzer test-IDs onto \texttt{CODE-*} rules; a few noisy IDs are deliberately left unmapped to suppress known false positives. \texttt{merge\_static\_layers} then unions regex and AST hits, the regex layer winning ties on the \((\texttt{rule\_id}, \texttt{line})\) key. Listing~\ref{lst:rule} shows a representative code rule.

\begin{lstlisting}[
  language=Rules,
  style=modelstyle,
  caption={Example deterministic code rule (YAML) grounding the LLM scanner.},
  label={lst:rule}
]
code_rules:
  - id: CODE-001
    severity: high
    pattern: 'os\.system\s*\(|shell\s*=\s*True'
    title: "Command-injection sink"
    rationale: "Unsanitized input reaching a shell"
    asset_category: "Material and Economic Resources"
\end{lstlisting}

\textbf{Detection: LLM refinement:} Each scanner then invokes its LLM with the deterministic hits injected as context and a strict instruction to \emph{refine, explain, or extend} them rather than invent new ones (paper 3's RAG-grounding pattern). \texttt{merge\_findings} keeps the deterministic finding whenever the LLM merely restates the same rule at the same line or evidence prefix, and admits an LLM finding only when it lands on a genuinely new locus. The source-code scanner adds a second ``what did you miss?'' critic pass aimed at semantic defects regexes cannot see: data races, TOCTOU, use-after-free, integer overflow. Its prompt is deliberately conservative (findings must map to a loaded rule):

\begin{promptbox}[Source-code scanner: LLM refinement pass]
You are a source-code security scanner. Flag only vulnerabilities that match a loaded rule: command/shell-injection sinks, weak credential crypto, hard-coded secrets, unsafe deserialization, missing input validation, and race conditions in privileged paths.\\[2pt]
For each finding return rule\_id, severity (low/medium/high/critical), a best-effort line number, a short evidence excerpt, and an explanation of why it is a vulnerability and which asset it threatens.\\[2pt]
Return strict JSON: \{"findings": [\{...\}]\}
\end{promptbox}

\textbf{Decision:} The decision agent aggregates all scanner findings into one merged set (the union of the three \texttt{AgentResult} finding lists) and reasons over it to choose a response tier (none/soft/moderate/hard), a confidence in $[0,1]$, the asset categories touched, and a justification. The tier is not the LLM's opinion alone: a deterministic, \emph{escalate-only} policy sits on top. First, signed cross-session memory is queried for similar past incidents, ranked by the Jaccard overlap of their rule-ID sets, and for a cross-session recommendation (a stronger tier whose remediation previously \emph{worked} at similarity $\geq 0.5$); such a precedent can raise, never lower, the tier. Second, if a prior round this run was refused or failed, \texttt{\_apply\_feedback\_escalation} bumps the tier at least one notch and nudges confidence. A malformed model reply fails safe to tier \texttt{none} at confidence 0.0. The full prompt is shown below.

\begin{promptbox}[Decision agent]
You are the central decision agent in a multi-agent cybersecurity defense system for an automotive/edge deployment.\\[2pt]
Aggregated findings from scanner agents: [findings]\\
Similar past incidents recalled from signed memory (paper-5 adaptive learning, precedent not gospel): [memory]\\
Cross-session recommendation (what worked before): [recommendation]\\
Feedback from the previous response attempt this run (empty on the first attempt): [feedback]\\
Allowed asset categories (paper-2 taxonomy): [assets]\\
Response tiers and thresholds: [tiers]\\[2pt]
Decide: (1) which asset categories are touched; (2) the recommended response tier; (3) an overall confidence in [0,1]; (4) a short justification.\\[2pt]
Return JSON: \{"summary": ..., "asset\_categories\_touched": [...], "recommended\_tier": "soft|moderate|hard|none", "confidence": <float>, "justification": ...\}
\end{promptbox}

\textbf{Action generation:} The action generator turns the decision into a multi-step remediation plan. A deterministic, rule-ID-keyed playbook is the floor; on top of it, an optional LLM tool-selector may \emph{add} steps from an MCP-style tool registry~\cite{mcp} in which every tool carries a schema, a declared side effect, and pre/post-conditions with step-level rollback. The prompt constrains the model to read-only/observe/verify tools and forbids destructive actions, and every proposed step is schema-validated and then re-checked by the same per-step guard as the playbook steps, so the LLM can strengthen but never weaken the plan.

\begin{promptbox}[Action generator: LLM tool selection]
You are the response planner for a defensive security action engine. A deterministic playbook has already chosen the core steps. You may propose ADDITIONAL tool calls that strengthen the response, preferring read-only/observe/verify tools that gather evidence or confirm the fix. Do NOT invent destructive actions.\\[2pt]
Decision summary: [summary] (tier=[tier], assets=[assets]).\\
Findings: [findings]\\
Available MCP tools (name [domain/side\_effect]: description): [catalogue]\\[2pt]
Return strict JSON: \{"tools": [\{"tool": <name>, "kind": "observe|audit|notify|contain", "params": \{...\}, "why": ...\}]\}. Return an empty list if the playbook suffices.
\end{promptbox}

\begin{figure*}[t]
    \centering
    \resizebox{\textwidth}{!}{%
    \begin{tikzpicture}
      \node[font=\scriptsize\bfseries, detblue,   rotate=90] at (-3.0,6)   {DETECTION};
      \node[font=\scriptsize\bfseries, decpurple, rotate=90] at (-3.0,1.4) {RESPONSE};
      \node[nio, text width=4.8cm, align=left] (inp) at (0,6)
        {{\scriptsize\bfseries INPUT: SOURCE CODE}\\[3pt]
         {\ttfamily\tiny def ping(host):\\ \ \ os.system("ping -c 1 "+host)}};
      \node[font=\scriptsize\bfseries, detblue] at (5.2,7.9) {Static analyzers};
      \node[ndet, text width=2.5cm] (reg) at (5.2,7.1) {\textbf{Regex Rules}\\ pattern floor};
      \node[ndet, text width=2.5cm] (ban) at (5.2,6.0) {\textbf{Bandit}\\ AST security lints};
      \node[ndet, text width=2.5cm] (scs) at (5.2,4.9) {\textbf{Source-Code Scanner}\\ LLM refinement};
      \node[nio, text width=2.8cm, align=left] (find) at (9.6,6)
        {{\scriptsize\bfseries FINDING}\\[3pt]
         {\ttfamily\scriptsize CODE-001}\\ Command-injection sink\\ line 2};
      \draw[fflow] (inp.east) -- (ban.west);
      \draw[fflow] (6.45,6.0) -- node[above, font=\tiny, black!60]{finding} (find.west);
      \node[font=\scriptsize, black!55] at (4.9,3.4) {the finding is handed to the response agents};
      \node[ndec, text width=3.6cm, align=left] (b1) at (0,1.4)
        {\textbf{Decision Agent}\\ weighs severity and past cases\\[3pt]
         \textbf{HIGH} severity, \textbf{92\%} confidence\\[2pt]
         {\footnotesize Unsanitized input reaches a shell command, remotely exploitable, so it warrants a hard fix.}};
      \node[ndec, text width=3.6cm, align=left] (b2) at (4.9,1.4)
        {\textbf{Action Generator}\\ plans the remediation\\[3pt]
         MCP tool calls:\\
         {\ttfamily\tiny read\_file("input.py")\\ quarantine("CODE-001")\\ open\_ticket("high")}};
      \node[ngrd, text width=2.8cm, align=left] (b3) at (9.3,1.4)
        {\textbf{Security Guard}\\ validates before run\\[4pt]
         {\color{grdgreen}\textbf{\checkmark\ Approved}}};
      \node[nio, text width=5.2cm, align=left] (b4) at (14.2,1.4)
        {{\scriptsize\bfseries OUTPUT: SAFE CODE}\\[3pt]
         {\ttfamily\tiny import ipaddress, shutil, subprocess\\ def ping(host):\\ \ ipaddress.ip\_address(host)\\ \ exe = shutil.which("ping")\\ \ subprocess.run([exe,"-c","1",host])}};
      \draw[fflow] (b1.east) -- (b2.west);
      \draw[fflow] (b2.east) -- (b3.west);
      \draw[fflow] (b3.east) -- (b4.west);
      \draw[fflow] (find.south) to[out=-90, in=90] ([xshift=9mm]b1.north);
    \end{tikzpicture}}
    \caption{A full worked example. An input artifact is scanned into static findings, annotated and merged, tiered by the decision agent, remediated by the action generator, and validated by the guard before the fix is committed.}
    \label{fig:walk}
\end{figure*}
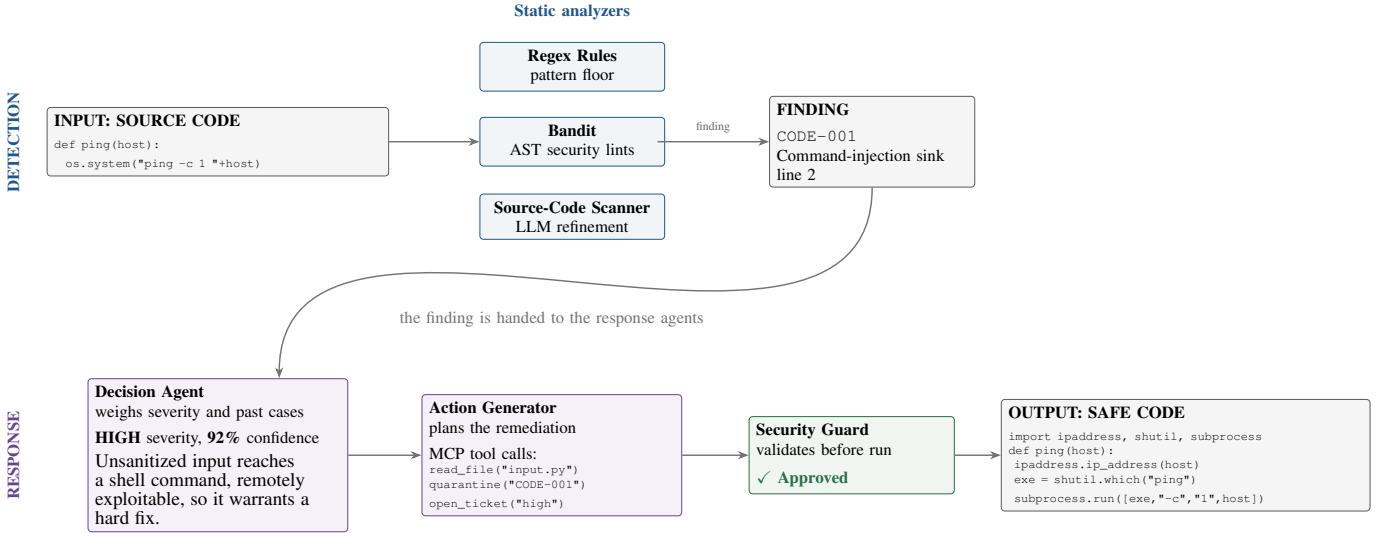

\textbf{Security guard:} Before any step executes, the \texttt{SecurityGuard} enforces paper-1's four contextual properties deterministically: \emph{task alignment} (the objective lies in the allowed set $O$), \emph{action alignment} (the action kind is compatible with the objective, via an auditable action$\rightarrow$objective map), \emph{source authorization} (every input that fed the action came from an authenticated source), and \emph{data isolation} (the action writes only to trusted destinations). Action alignment is additionally gated by the independent $H_a$ oracle, a \emph{separate} LLM call whose only job is a 0/1 alignment verdict, memoized per \((\text{action},\text{objective},\text{context})\). It can only tighten the deterministic floor (\(\text{action\_ok}=\text{static\_map\_ok}\wedge H_a\text{\_ok}\)); keeping it separate means a prompt injection that subverts the scanners or decision agent still faces an untainted judge. The guard also evaluates a whole plan for compositional and temporal safety (paper 1 \S7). If a step is refused or a plan fails, the failure is fed back and the loop re-plans with an escalated tier; if approved, the change is committed and the outcome is written, HMAC-signed and source-tagged, to the incident store for future recall. Fig.~\ref{fig:walk} traces one scenario through the full pipeline, from raw input to committed fix.

\begin{promptbox}[Security guard: $H_a$ action-alignment oracle]
You are the action-alignment oracle $H_a$ from the contextual-security framework (paper 1). Your ONLY job: decide whether the proposed ACTION genuinely serves the declared OBJECTIVE in the given CONTEXT. An action can be individually valid yet still NOT serve the objective (e.g., a privileged change requested under a read-only objective, or acting on a trusted asset).\\[2pt]
OBJECTIVE: [objective]\\
ACTION: [action]\\
CONTEXT: [context]\\[2pt]
Return strict JSON: \{"serves": true|false, "reason": ...\}
\end{promptbox}

\subsection{Attack Workflow}
To measure defense quality on identical scenarios, CyberLLM includes a symmetric red-team pipeline whose stages mirror the defense agents one-for-one and reuse the same rule base.

\textbf{Recon and exploiters.} A system-information discovery agent does \emph{passive} reconnaissance only, materializing the AgentHeLLM actor/datasource graph as a structured \texttt{SystemInstance}; no packets are sent. Two exploiter agents then reuse the \emph{same} static-analysis hits as the defense scanners (finding a bug is identical in both modes; only intent differs): the source-code exploiter turns each finding into an exploit plan tagged with the CWE and which of paper~1's four properties it violates, and a deployment exploiter derives steps from the topology graph.

\textbf{Orchestration and ethics guard.} An attack orchestrator, the mirror of the decision agent, chains one to three findings against a paper~2 asset target and lays out a bi-level \emph{poison path} (payload delivery) and \emph{trigger path} (victim consumption); a malformed reply falls back to chaining the highest-severity findings. An attack generator then emits one non-destructive, canary-only script per step (shell-injection, pickle RCE, path-traversal, SQL-injection, \texttt{alg=none} JWT, agent-to-agent poison, CAN frame), returned as text and never executed. Every plan first passes an ethics guard, the offensive analogue of the \texttt{SecurityGuard}, checking scenario scope, explicit target authorization, and the absence of destructive patterns, and redacting the script on any failure. Outcomes are written to the same HMAC-signed store tagged as \texttt{attack} mode, so both sides are directly comparable and blind tests can be scored without leaking labels.

\subsection{Cross-cutting Mechanisms}
Three mechanisms span the pipeline. \emph{Signed memory} is a stdlib SQLite store; every write is HMAC-signed with a per-deployment key and tagged with the writing source, and reads verify the signature and source before a row can influence a decision, defending the learning loop against poisoning. \emph{Asset tagging} maps each high-severity finding to one of the seven human-centric categories, giving downstream steps human-impact context. \emph{Provenance} is enforced at ingestion so that source-authorization checks are always decidable.

\section{Experiments and Evaluation}
\label{sec:eval}

\subsection{Goal and Environment}
The evaluation targets the project's core question: how much of a realistic automotive attack surface can the pipeline surface from source code alone, and at what precision. We therefore benchmark detection against an independent, ground-truthed corpus of original automotive ECU code rather than the illustrative single-file scenarios used above. The corpus (\texttt{benchmark/automotive/}) comprises nine vulnerable modules written in C, C++, and Rust and modelled on real ECU subsystems, plus two clean negative controls. The nine modules are a CAN-to-Ethernet gateway, a UDS (ISO~14229) security-access handler, a verified-boot loader, an OTA firmware-update agent, an MQTT telemetry publisher, an ADAS sensor-fusion buffer, a CAN-FD parser with a C foreign-function interface, an in-vehicle key store, and a local fleet vehicle API. The code carries no hint comments; every flaw is recorded only in a separate \texttt{ground\_truth.yaml}, so the corpus was written to test the detector, not to flatter it. In total it seeds \textbf{47 layered, non-obvious vulnerabilities} spanning memory safety and concurrency (CWE-787/125/120/190/191/416/476/362), injection (CWE-78/89/22/134), cryptography and secrets (CWE-798/330/327/208/916/347/295), authorization (CWE-287/862/1390), and reachable-panic denial of service (CWE-248). We evaluate two operating points: a \emph{Fast} mode (the deterministic layer only: regex floor plus Bandit~\cite{bandit}/Semgrep~\cite{semgrep} where present, fully offline) and a \emph{Deep} mode (Fast plus the live LLM refinement and completeness critic passes, run with GPT-4o accessed via Azure OpenAI service~\cite{gpt4o}).

\subsection{Metrics}
A labelled vulnerability counts as detected when the scanner emits a finding whose line lies within a $\pm 3$-line window of the labelled line or range; matching is greedy and one-to-one, and a finding's \texttt{rule\_id} need not equal the CWE (CWE and category are recorded for reporting, not to penalise naming differences). We report \emph{coverage/recall} (detected $\div$ total labelled), broken down by language, severity, and vulnerability category. \emph{Precision} is measured on the clean control files, where any finding is a true false positive: $\text{precision}=\text{TP}/(\text{TP}+\text{FP}_{\text{clean}})$. We also report the harmonic-mean \emph{F1}. Findings on a vulnerable file that match no label are reported as extra (unlabeled) rather than scored as errors, because the label set is a must-have floor, not an exhaustive oracle. The harness is static-first, so mean-time-to-detect is N/A by design.

\subsection{Results}
Table~\ref{tab:benchmark} reports coverage, precision, recall, and F1 for both modes, with a per-language breakdown. The deterministic Fast layer covers 34.0\% (16/47) of the labelled vulnerabilities at perfect precision, giving a zero-cost, fully offline floor. Enabling the LLM refinement and completeness passes (Deep) raises coverage to $\approx 70\%$ (33/47) and F1 to 0.83 while holding precision at 1.000: the clean controls produce zero findings in every run because the completeness critic is instructed to treat defensive patterns (parameterized queries, bounds-checked copies, constant-time compares, checked \texttt{Result}/\texttt{Option} handling) as safe and to report only high-confidence defects. The gain is concentrated in exactly the classes a regex floor cannot see: medium-severity logic and DoS defects (subtle races, off-by-one, use-after-free, reachable panics) rise from 3/11 to roughly 8/11 once the second pass is on.

\begin{table}[t]
\caption{Detection on the automotive C/C++/Rust benchmark (47 labelled vulnerabilities, 9 modules, 2 clean controls), location-based matching at $\pm 3$ lines. Precision is measured on the clean controls.}
\label{tab:benchmark}
\centering
\footnotesize
\begin{tabular}{lcc}
\toprule
\textbf{Metric} & \textbf{Fast} & \textbf{Deep} \\
                & \textbf{(offline)} & \textbf{(LLM + critic)} \\
\midrule
Coverage        & 34.0\% (16/47) & $\sim$70\% (33/47) \\
Precision       & 1.000          & 1.000 \\
Recall          & 0.34           & 0.70 \\
F1              & 0.51           & 0.83 \\
\midrule
C (15 vulns)    & 3/15           & 9/15 \\
C++ (16 vulns)  & 6/16           & 12/16 \\
Rust (16 vulns) & 7/16           & 12/16 \\
\bottomrule
\end{tabular}
\end{table}

The improvement was incremental and reproducible. Starting from a Python-centric first cut at 25.5\% coverage and 0.41 F1, adding multi-language ingestion (C/C++/Rust source and headers) and native-language rules lifted the offline floor and anchored the LLM to 55.3\% coverage and 0.71 F1; the completeness second pass then brought Deep mode to $\approx 70\%$ coverage and $\approx 0.83$ F1. Every native rule was verified not to fire on the clean controls, so each step preserved the 1.000 precision. As LLM is non-deterministic, Deep coverage varies a few points run to run, while Fast mode is reproducible.

\subsection{Discussion}
Three points follow from these numbers. First, the split between the deterministic floor (34.0\% at 1.000 precision) and the LLM-augmented detector ($\approx 70\%$ at 1.000 precision) quantifies exactly what the LLM contributes when it is grounded and bounded rather than trusted blindly: it roughly doubles recall without introducing a single false positive on benign code, which is the central design claim of the layered detector. Second, the residual misses are informative rather than incidental: they concentrate in medium-severity logic and DoS defects (races, off-by-one, NULL dereference, reachable panics) across all three languages, marking the frontier for dynamic analysis and richer rule grounding. Third, and most important for a safety-critical deployment, the detection numbers are only the input to a guarded response: because the guard is escalate-only, provenance is mandatory, and the $H_a$ oracle is an independent model call that can only tighten the verdict, an adversarial input that subverts a scanner still cannot drive an unsafe action to commit, so recall gains never come at the cost of the safety envelope.

\textbf{Availability.} The full implementation, including a web-based user interface for interactively exercising both the defense and attack pipelines as well as the benchmarking harness used in this evaluation, is publicly available at \url{https://github.com/Jeddou10/CyberLLM}.

\section{Conclusion}
\label{sec:concl}

This paper presented a multi-agent LLM framework that couples autonomous, cross-modal vulnerability detection with a formally-guarded, tiered response for automotive cybersecurity. By grounding LLM scanners with deterministic analyzers, ratcheting response decisions with tamper-evident memory, and enforcing four contextual security properties plus an independent action-alignment oracle at runtime, the system covers about 70\% of the labeled vulnerabilities on an independent C/C++/Rust automotive benchmark at 1.000 precision, roughly doubling the deterministic floor, and never commits an unsafe action without approval. A symmetric attack pipeline enables both sides to be measured on identical scenarios.

Future work will integrate live MCP tool servers (e.g., \texttt{nmap}, \texttt{tcpdump}, CAN replay) for active detection, add retrieval-augmented grounding over CVE and standards corpora to further curb hallucination, handle large models and code bases via Model Context Protocol and RAG, and validate the pipeline against high-fidelity SDV testbenches with autonomous driving capabilities and integrated with simulation environment, such as CARLA \cite{lebioda2025requirements}. We also plan a broader evaluation against agentic-security benchmarks to characterize robustness under adaptive adversaries.


\vspace{12pt}


\begin{thebibliography}{00}

\bibitem{siu2026framework}
K.~Siu, Z.~He, C.~Montgomery, Y.~Wang, Z.~Gong, Y.~Wang, and D.~Song, ``A Framework for Formalizing LLM Agent Security,'' \emph{arXiv preprint arXiv:2603.19469}, 2026.

\bibitem{stappen2026agent2agent}
L.~Stappen, D.~Turan, G.~Hagerer, and G.~Groh, ``Agent2Agent Threats in Safety-Critical LLM Assistants: A Human-Centric Taxonomy,'' \emph{arXiv preprint arXiv:2602.05877}, 2026.

\bibitem{petrovic2026fusa}
N.~Petrovic, V.~Zolfaghari, F.~Pan, and A.~Knoll, ``LLM-Empowered Functional Safety and Security by Design in Automotive Systems,'' \emph{arXiv preprint arXiv:2601.02215}, 2026.

\bibitem{harer2024multiagent}
F.~H\"arer, ``Multi-Agent LLM Systems: A Schema-Driven Execution Framework with Cybersecurity Case Studies,'' \emph{arXiv preprint}, 2024.

\bibitem{collins2026orchestrated}
J.~Collins, A.~James, and R.~Thomas, ``LLM-Orchestrated Multi-Agent Coordination for Adaptive Cyber Defense at the Edge,'' 2026.

\bibitem{iso21434}
International Organization for Standardization, ``ISO/SAE 21434: Road Vehicles, Cybersecurity Engineering,'' 2021.

\bibitem{pearce2023repair}
H.~Pearce, B.~Ahmad, B.~Tan, B.~Dolan-Gavitt, and R.~Karri, ``Examining Zero-Shot Vulnerability Repair with Large Language Models,'' in \emph{IEEE Symposium on Security and Privacy (S\&P)}, 2023.

\bibitem{ji2023hallucination}
Z.~Ji, N.~Lee, R.~Frieske, T.~Yu, D.~Su, Y.~Xu, E.~Ishii, Y.~J. Bang, A.~Madotto, and P.~Fung, ``Survey of Hallucination in Natural Language Generation,'' \emph{ACM Computing Surveys}, vol.~55, no.~12, pp.~1--38, 2023.

\bibitem{greshake2023injection}
K.~Greshake, S.~Abdelnabi, S.~Mishra, C.~Endres, T.~Holz, and M.~Fritz, ``Not What You've Signed Up For: Compromising Real-World LLM-Integrated Applications with Indirect Prompt Injection,'' in \emph{ACM Workshop on Artificial Intelligence and Security (AISec)}, 2023.

\bibitem{gpt4o}
OpenAI, ``GPT-4o System Card,'' \emph{arXiv preprint arXiv:2410.21276}, 2024.

\bibitem{owaspagentic}
OWASP, ``OWASP Top 10 for Large Language Model Applications,'' 2025. [Online]. Available: \url{https://owasp.org/www-project-top-10-for-large-language-model-applications/}

\bibitem{bandit}
PyCQA, ``Bandit: A Security Linter for Python,'' 2024. [Online]. Available: \url{https://bandit.readthedocs.io/}

\bibitem{semgrep}
Semgrep, ``Semgrep: Lightweight Static Analysis for Many Languages,'' 2024. [Online]. Available: \url{https://semgrep.dev/}

\bibitem{mcp}
Anthropic, ``Model Context Protocol (MCP) Specification,'' 2024. [Online]. Available: \url{https://modelcontextprotocol.io/}

\bibitem{lebioda2025requirements}
K. Lebioda, N. Petrovic, F. Pan, V. Zolfaghari, A. Schamschurko, and A. Knoll, ``Are Requirements Really All You Need? Using LLMs to Generate Configuration Code: A Case Study in Automotive Simulations,'' \emph{IEEE Access}, vol. 13, pp. 145115--145126, 2025. [Online]. Available: \url{https://doi.org/10.1109/ACCESS.2025.3597748}

\end{thebibliography}
\end{document}